\documentclass[namedreferences]{kluwer}
\usepackage[]{graphicx}
\begin{document}

\def\ffam {\hbox{$\,.\!\!^{\prime}$}}
\def\ffas {\hbox{$\,.\!\!^{\prime\prime}$}}
\def\ffM {\hbox{$\,.\!\!^{\rm M}$}}
\def\ffm {\hbox{$\,.\!\!^{\rm m}$}}
\def \la{\mathrel{\mathchoice   {\vcenter{\offinterlineskip\halign{\hfil
$\displaystyle##$\hfil\cr<\cr\sim\cr}}}
{\vcenter{\offinterlineskip\halign{\hfil$\textstyle##$\hfil\cr
<\cr\sim\cr}}}
{\vcenter{\offinterlineskip\halign{\hfil$\scriptstyle##$\hfil\cr
<\cr\sim\cr}}}
{\vcenter{\offinterlineskip\halign{\hfil$\scriptscriptstyle##$\hfil\cr
<\cr\sim\cr}}}}}
\def \ga{\mathrel{\mathchoice   {\vcenter{\offinterlineskip\halign{\hfil
$\displaystyle##$\hfil\cr>\cr\sim\cr}}}
{\vcenter{\offinterlineskip\halign{\hfil$\textstyle##$\hfil\cr
>\cr\sim\cr}}}
{\vcenter{\offinterlineskip\halign{\hfil$\scriptstyle##$\hfil\cr
>\cr\sim\cr}}}
{\vcenter{\offinterlineskip\halign{\hfil$\scriptscriptstyle##$\hfil\cr
>\cr\sim\cr}}}}}

\begin{article}
\begin{opening}
\title{H$_2$O megamasers: Accretion disks, jet interaction, outflows or massive star formation?}    

\author{C.   \surname{Henkel$^{1}$}}
\author{J.A. \surname{Braatz$^{2}$}}
\author{A.   \surname{Tarchi$^{3,4}$}}
\author{A.B. \surname{Peck$^{5}$}}
\author{N.M. \surname{Nagar$^{6}$}}
\author{L.J. \surname{Greenhill$^{7}$}}
\author{M.   \surname{Wang$^{8}$}}
\author{Y.   \surname{Hagiwara$^{9}$}}
\institute{$^{1}$Max-Planck-Institut f{\"u}r Radioastronomie, Auf dem H{\"u}gel 69, D--53121 Bonn, Germany} 
\institute{$^{2}$NRAO, P.O. Box 2, Green Bank, WV 24944, USA}
\institute{$^{3}$Istituto di Radioastronomia, CNR, Via Gobetti 101, I--40129 Bologna, Italy}
\institute{$^{4}$INAF-Osservatorio Astronomico di Cagliari, Loc. Poggio dei Pini, Strada 54, I--09012 Capoterra (CA), Italy}
\institute{$^{5}$Harvard-Smithsonian Center for Astrophysics, SAO/SMA Project, 645 N. A'ohoku Pl., Hilo, HI 96720, USA}
\institute{$^{6}$Kapteyn Instituut, Postbus 800, NL--9700 AV Groningen, The Netherlands}
\institute{$^{7}$Harvard-Smithsonian Center for Astrophysics, 60 Garden St., Cambridge MA 02138, USA}
\institute{$^{8}$Purple Mountain Observatory, Chinese Academy of Sciences, 210008 Nanjing, China}
\institute{$^{9}$ASTRON/Westerbork Radio Observatory, P.O. Box 2, NL--Dwingeloo 7990 AA, The Netherlands}



\runningtitle{H$_2$O megamasers }
\runningauthor{Henkel et al.}

\begin{ao}
Max-Planck-Institut f{\"u}r Radioastronomie\\
Auf dem H{\"u}gel 69\\
D-53121 Bonn\\
Germany
\end{ao} 


\begin{abstract}

The 25 years following the serendipitous discovery of megamasers have seen tremendous progress 
in the study of luminous extragalactic H$_2$O emission. Single-dish monitoring and high resolution 
interferometry have been used to identify sites of massive star formation, to study the interaction
of nuclear jets with dense molecular gas and to investigate the circumnuclear environment of active galactic 
nuclei (AGN). Accretion disks with radii of 0.1--3\,pc were mapped and masses of nuclear engines of order 
10$^{6}$--10$^{8}$\,M$_{\odot}$ were determined. So far, $\sim$50 extragalactic H$_2$O maser sources have 
been detected, but few have been studied in detail.
\end{abstract}




\end{opening}

\section{Introduction}

To date, maser lines of five molecular species, those of CH, OH, H$_2$O, SiO, and H$_2$CO, have been detected in 
extragalactic space. While the number of observed OH masers, $\sim$100, is largest, the greatest emphasis of astrophysical 
research is focused on H$_2$O. This is a consequence of the fact that the 6$_{16}$--5$_{23}$ line of interstellar water 
vapor at 22.235\,GHz ($\lambda$$\sim$1.3\,cm), first detected towards Orion-KL, Sgr\,B2 and W\,49 (\opencite{cheung69}) 
and, outside the Galaxy, towards the nearby spiral M\,33 (\opencite{churchwell77}), can be efficiently used to pinpoint 
sites of massive star formation and to elucidate the properties of active galactic nuclei (AGN). The line traces dense 
($n$(H$_2$)$\ga$10$^{7}$\,cm$^{-3}$) warm ($T_{\rm kin}$$\ga$400\,K) molecular gas. Characteristic properties are 
enormous brightness temperatures ($\sim$10$^{12}$\,K has been measured), small sizes of individual hotspots 
($\la$10$^{14}$\,cm in galactic sources) and narrow linewidths (typically a few km\,s$^{-1}$) that make these masers 
to ideal probes of the structure and dynamics of the gas in which they reside. Because the apparent isotropic luminosity 
of these masers can be truly outstanding, reaching 10$^{3-4}$ L$_{\odot}$ (up to 10$^{53}$ photons/s) in the most extreme 
cases, H$_2$O masers can be observed out to fairly large distances (up to c$z$$\sim$17700\,km\,s$^{-1}$; \opencite{tarchi03}).

\section{Kilomasers related to star formation}

In the Galaxy, H$_2$O masers from star forming regions have isotropic luminosities of order $L_{\rm H_2O}$ $\la$
10$^{-2}$\,L$_{\odot}$, with the notable exception of W\,49 ($L_{\rm H_2O}$ $\sim$ 1\,L$_{\odot}$). The more
luminous of these sources consist of a number of hotspots (see Brand et al., this volume). In nearby galaxies, 
masers with similar luminosities, the so-called kilomasers, are also found with luminosities up to a few L$_{\odot}$.
Such masers are important to pinpoint sites of massive star formation (e.g. \opencite{tarchi02}) and to estimate 
distances on a purely geometric basis, comparing radial velocity and proper motion dispersions in groups of maser 
spots (\opencite{greenhill93}; \opencite{argon04}). Making use of known rotational properties, proper motions can 
also be used to determine three dimensional velocity vectors of entire galaxies (e.g. \opencite{brunthaler02}), 
thus providing information on the content and distribution of mass inside the Local Group.

\begin{figure}
\centerline{\includegraphics[width=8cm]{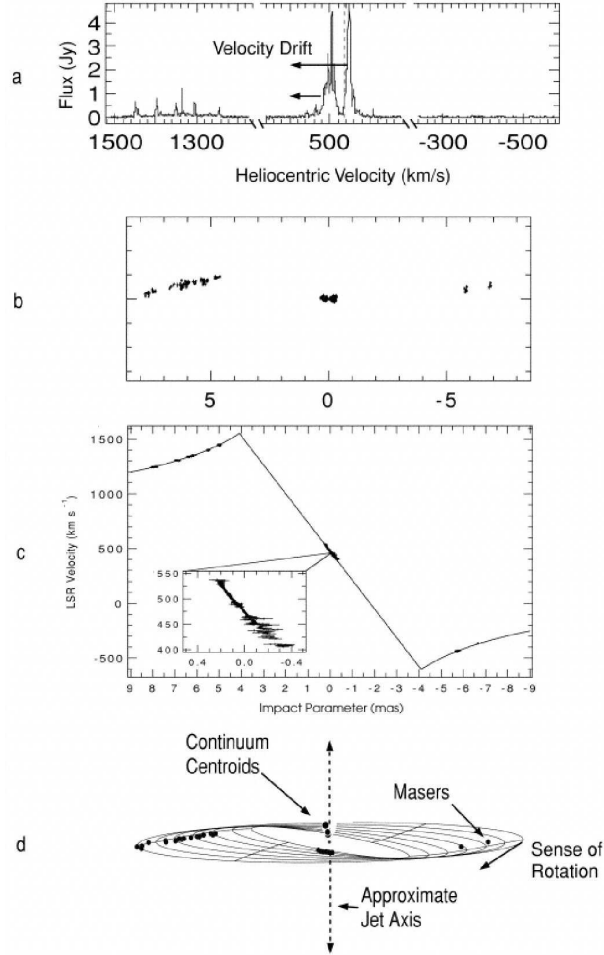}}
\caption{Overview of the NGC\,4258 system: a) Typical spectrum, b) VLBA (Very Long Baseline Array) map, c) velocity 
versus impact parameter with Keplerian rotation curve fit, and d) warped disk with VLBA map (from Bragg et al. 2000).}
\label{ngc4258a}
\end{figure}

\section{Accretion disk megamasers}

Megamasers, with $L_{\rm H_2O}$$\ga$20\,L$_{\odot}$, were first observed towards NGC\,4945 (\opencite{dossantos79}) 
and then in NGC\,1068, NGC\,3079, NGC\,4258 and the Circinus galaxy (\opencite{gardner82}; \opencite{claussen84}; 
\opencite{henkel84}; \opencite{haschick85}). Two of these masers, those of NGC\,1068 and NGC\,4258, were soon found 
to be located in the innermost few parsecs of their parent galaxies (\opencite{claussen86}). Selection criteria to 
find more such masers as well as an interpretation of the molecular line emission in terms of properties characterizing 
the nuclear environment of AGN remained, however, elusive for a full decade. A large survey including $\sim$360 Seyfert 
and LINER galaxies (\opencite{braatz96}) finally led to the identification of another 10 megamasers, thus tripling 
the number of known sources and permitting, for the first time, a statistical analysis. All of these masers were found 
to be associated with Seyfert 2 or LINER nuclei (\opencite{braatz97}). Adopting the so-called unified scheme in which 
Seyfert 1s and 2s are identical except for angle of view, H$_2$O megamaser activity then is related to the large 
line-of-sight column densities expected when the nuclear tori are viewed edge-on. At least some LINER galaxies also 
contain large column densities of warm dense molecular gas as well as an AGN.

\begin{figure}
\centerline{\includegraphics[width=7.5cm]{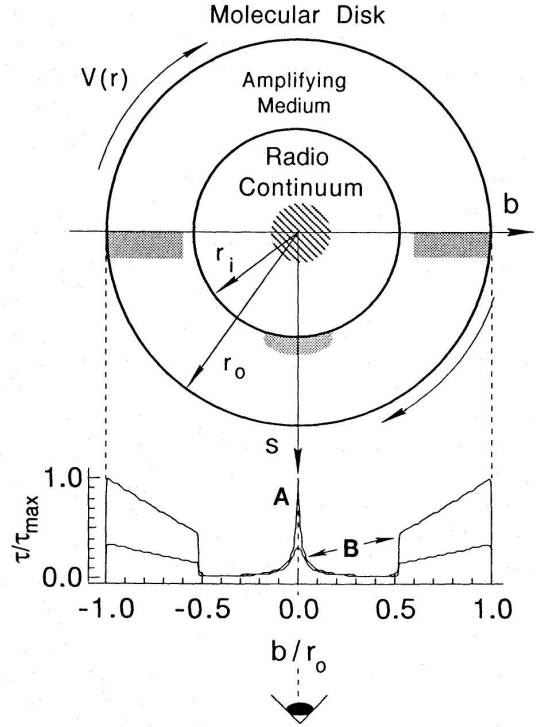}}
\caption{The Keplerian disk of NGC\,4258 seen from above. The normalized optical depth along various lines of sight is 
shown for the plane of the disk, where $b$ is the impact parameter. The shaded areas denote lines-of-sight with detected
maser emission. $r_{\rm i}$ and $r_{\rm o}$ are the inner and outer radii of the masing disk, $s$ marks the distance
from its midline (from Greenhill et al. 1995a).}
\label{ngc4258b}
\end{figure}

More detailed studies of the megamaser in NGC\,4258 not only confirmed the presence of $>$10 systemic velocity
features at a given time but also led to the detection of two additional H$_2$O velocity components, $\pm$1000\,km\,s$^{-1}$ 
off the systemic velocity (\opencite{nakai93}). The systemic features show a secular drift, with radial velocities 
increasing by d$V_{\rm s}$/d$t$$\sim$10\,km\,s$^{-1}$\,yr$^{-1}$ (\opencite{haschick90}; \opencite{haschick94}; 
\opencite{greenhill95a}; \opencite{nakai95}). Very Long Baseline Interferometry (VLBI) reveals the presence of a 
warped edge-on Keplerian disk of $\sim$0.5\,pc diameter (\opencite{greenhill95b}; \opencite{miyoshi95}; 
\opencite{herrnstein98a}, \opencite{herrnstein99}), as shown in Fig.\,\ref{ngc4258a}. Its geometry reflects the velocity 
coherence of the differentially rotating gas along the lines-of-sight towards its front side, its back side and its 
tangentially viewed regions (Fig.\,\ref{ngc4258b}) as well as possible amplification of the radio continuum of the 
northern jet by the systemic features. 

The positive velocity drift of the systemic component is readily explained by centripetal acceleration, implying that 
the features arise from the front and not from the back side of the masing disk (the latter would result in a negative 
drift). The `high velocity features' originate from those parts of the disk seen tangentially. Applying the virial 
theorem yields 
\begin{equation}
M_{\rm core} = 1.12 \left[\frac{V_{\rm rot}}{\rm km\,s^{-1}}\right]^2 \left[\frac{R}{\rm mas}\right] 
             \left[\frac{D}{\rm Mpc}\right] {\rm M_{\odot}},
\end{equation}
with $M_{\rm core}$ being the mass enclosed by the Keplerian disk, $V_{\rm rot}$ denoting its rotational velocity at 
angular radius $R$, and $D$ representing the distance to the galaxy. The VLBI maps allow us to directly measure 
$V_{\rm rot}$ and $R$ for various values of $R$. From the Keplerian rotation curve we then obtain
\begin{equation}
{\rm C_1} = \left[\frac{V_{\rm rot}}{\rm km\,s^{-1}}\right] \left[\frac{R}{\rm mas}\right]^{1/2}.
\end{equation}
In addition, the observed E-W velocity gradient of the systemic features (Fig.\,1c) provides
\begin{equation}
{\rm C_2} = \left[\frac{V_{\rm rot}}{\rm km\,s^{-1}}\right] \left[\frac{R}{\rm mas}\right]^{-1}.
\end{equation}
$C_1$/$C_2$ = $R^{3/2}$ then gives the angular radius $R_{\rm s}$ of the systemic features as viewed from a direction 
in the plane of the disk, but perpendicular to the line-of-sight. The result, $R_{\rm s}$$\sim$\,4.1\,mas, implies 
that the features are localized toward the inner edge of the disk. The distance to the disk, needed to estimate 
$M_{\rm core}$ in Eq.\,(1), is determined by measuring the centripetal 
acceleration
\begin{equation}
{\rm d}V_{\rm s}/{\rm d}t = \frac{V_{\rm rot,s}^2}{r_{\rm s}} = 9.3\pm 0.3\,{\rm km\,s^{-1}\,yr^{-1}}.
\end{equation}
With d$V_{\rm s}$/d$t$ and the rotational velocity $V_{\rm rot,s}$ of the systemic components known, the linear scale 
$r_{\rm s}$ can be compared with the angular scale, $R_{\rm s}$, providing a measure of the distance. The proper 
motion of the individual systemic maser spots, 31.5$\pm$1.0\,$\mu$as\,yr$^{-1}$, allows, with $V_{\rm rot,s}$ being 
known, a check of this distance estimate. The results of the two methods agree. $D$=7.2$\pm$0.5\,Mpc and 
$M_{\rm core}$ = (3.9$\pm$0.3)$\times$10$^{7}$\,M$_{\odot}$ within 0.14\,pc (\opencite{herrnstein99}).

\begin{figure}
\centerline{\includegraphics[width=6cm]{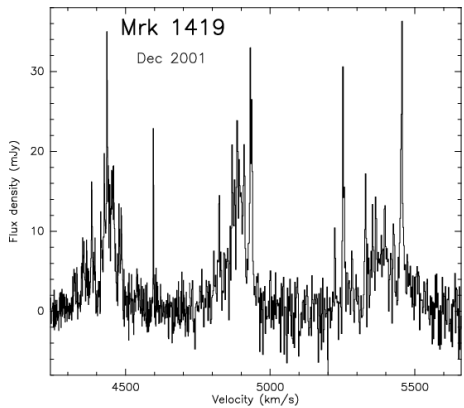}}
\caption{22\,GHz H$_2$O spectrum of Mrk\,1419 (NGC\,2960), taken with the 100-m telescope at Effelsberg 
(from Henkel et al. 2002) and showing the characteristic systemic (center), red- and blue-shifted groups of 
maser features.}
\label{mrk1419}
\end{figure}

So far, all H$_2$O megamasers studied interferometrically arise from the innermost parsecs of their parent galaxy 
and all appear to be associated with AGN. Searching for NGC\,4258-like targets showing three groups of H$_2$O
features, one systemic, one red- and one blue-shifted, has become the Holy Grail of recent maser surveys. Such a
configuration requires not only the presence of very dense molecular gas, but also a suitable viewpoint, with the
nuclear disk seen almost edge-on. Therefore such sources are rarely found. Nevertheless, the number of detected 
sources falling into this category is steadily rising (see e.g. Fig.\,\ref{mrk1419}). Most notable are IC\,2560 
(2900\,km\,s$^{-1}$; \opencite{ishihara01}) and Mrk\,1419 (4900\,km\,s$^{-1}$; \opencite{henkel02}). Very 
recently, four additional sources were identified, one at 7900\,km\,s$^{-1}$ (\opencite{braatz04}) and three at 
$>$10000\,km\,s$^{-1}$ (Greenhill et al., in preparation). Assuming that the model valid for NGC\,4258 is generally 
applicable, single dish observations determining the drift of the systemic components (d$V_{\rm s}$/d$t$) and the 
velocity offsets to the non-systemic components ($V_{\rm rot}$) are sufficient for a rough estimate of the enclosed 
mass and linear scale of the disk. Thus it is possible to estimate angular scales of order 1\,mas observing with 
resolutions of 35--40$''$. So far, all measured centripetal accelerations are positive. The most plausible explanations 
are obscuration of the back side, perhaps by free-free absorption, and a lack of background radio continuum emission 
that could be amplified.

Maser dynamical masses ($M_{\rm core}$) for a large sample of galaxies have the potential to establish the slope of 
the $M_{\rm core}$--$\sigma$ relation and its intrinsic scatter (e.g. \opencite{ferrarese00}), with uncertainties 
being dominated by the stellar velocity dispersion $\sigma$. Observing the three dimensional structure of a sample of 
accretion disks may provide strong constraints on heating (possibly irradiation by the nuclear X-ray source) and 
warping mechanisms (radiative torques have been proposed) and on the stability of these structures (e.g. 
\opencite{neufeld95}). The thickness of the respective accretion disk is another important parameter 
(e.g. \opencite{moran99}), being crucial to calculations of accretion rate and identification of accretion modes 
(e.g. advective, convective, viscous). Determining the mass of the nuclear source, its distance, Eddington 
luminosity and accretion efficiency, the rate and mode of the nuclear accretion flow, the size and geometry of the 
nuclear disk or torus, the parent galaxy's deviation from the Hubble flow and calibrations of optical or near infrared 
indicators of distance with geometrically obtained values (see e.g. Eqs.\,1--4) are all attractive goals. 

In addition to the sources introduced above, there are a large number of targets exhibiting somewhat less regular 
spectra. These may either arise from tori that are unstable to fragmentation and star formation or from sources that 
combine an edge-on nuclear accretion disk with maser components of different nature that will be discussed below. Within 
this context, the most thoroughly studied sources are NGC\,1068, NGC\,3079 and the Circinus galaxy (\opencite{gallimore96}; 
\opencite{greenhill96}, \opencite{greenhill97}, \opencite{trotter98}; \opencite{hagiwara02}; \opencite{greenhill03}; 
\opencite{kondratko04}).

\section{Jet megamasers}

There are also sources in which at least a part of the H$_2$O emission is believed to be the result of an interaction
between the nuclear radio jet and an encroaching molecular cloud. The first such source was NGC\,1068, where not only 
the three groups of H$_2$O components from the accretion disk are seen (\opencite{greenhill96}), but where a fourth 
component is also detected. This originates from a region 0\ffas3 ($\sim$30\,pc) downstream, where the radio jet bends 
(\opencite{gallimore96}; \opencite{gallimore01}). To maximize detection rates of jet-maser sources, galaxies should be 
selected that either show evidence for interaction between the radio jet and clouds in the narrow line region or that 
show a face-on ($i$$<$35$^{\circ}$) large scale galaxy disk and extended (up to 100\,pc) radio structures, indicating 
that both the disk and the radio jet are fairly close to the plane of the sky. 

\begin{figure}
\centerline{\includegraphics[width=8cm]{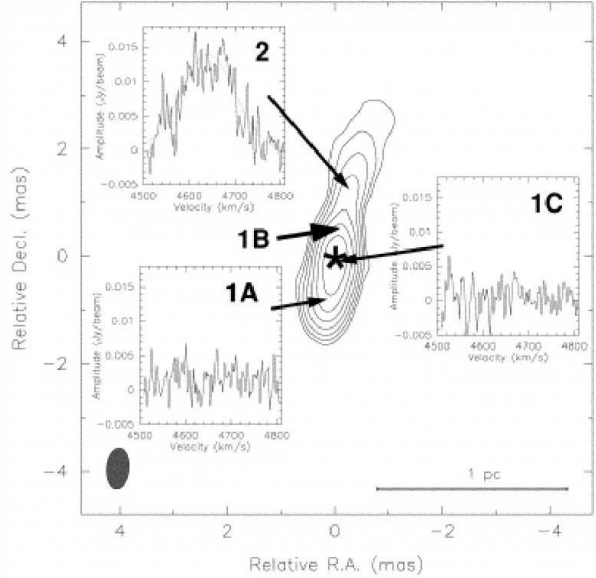}}
\caption{VLBA 22\,GHz continuum and H$_2$O spectra towards Mrk\,348 (NGC\,262; see Peck et al. 2003). H$_2$O emission is only 
detected towards the northern jet.}
\label{mrk348}
\end{figure}

Toward the elliptical galaxy NGC\,1052 the H$_2$O maser features are located along the line-of-sight to the south-western 
nuclear jet (\opencite{claussen98}). In Mrk\,348, a spiral galaxy with a particularly strong nuclear radio component, 
the megamaser appears to be associated with the northern jet (see Fig.\,\ref{mrk348}; \opencite{peck03}). The intensity 
of the line emission is correlated with the continuum flux. The high linewidth ($\sim$130\,km\,s$^{-1}$) on small spatial 
scales ($<$0.25\,pc) and the rapid variability indicate that the H$_2$O emission is more likely to arise from a shocked 
region at the interface between the energetic jet and the ambient molecular gas than as a result of amplification of the 
continuum jet by molecular clouds along the line-of-sight. The line emission, red-shifted by 130\,km\,s$^{-1}$ with 
respect to the systemic velocity, may arise from gas being entrained by the receding jet. The close temporal correlation 
between the flaring activity of the maser and the continuum further suggests that the H$_2$O and continuum hotspots are 
nearly equidistant from the central engine and may be different manifestations of the same event.

\section{New developments}

Recently, optically detected large scale outflows were proposed to induce shocks that might trigger H$_2$O megamaser 
emission (\opencite{schulz03}). Independently, it was found that in the Circinus galaxy the H$_2$O megamaser 
does not only trace a circumnuclear disk, but that there are also features associated with a wide angle outflow 
(\opencite{greenhill03}). H$_2$O maser emission traces this outflow out to 1\,pc from the central engine with 
line-of-sight velocities up to $\pm$160\,km\,s$^{-1}$ w.r.t. systemic. The outflowing wind is observed in those 
regions around the circumnuclear disk that are not shadowed by its warp. The position angles of the edges of the 
outflow correspond to those of the outflow and ionization cones observed at radio and optical wavelengths on much 
larger scales (\opencite{veilleux97}; \opencite{curran99}). Thus in addition to accretion disk and jet megamasers, 
strong maser emission can also be associated with large scale nuclear outflows. Yet another kind of H$_2$O megamaser 
might be associated with massive star formation. So far there is not yet a confirmed source. Towards Arp\,299, an 
interacting pair of galaxies, luminous H$_2$O megamaser emission was detected that might be associated with star 
formation in the overlapping region between its main members NGC\,3690 and IC\,694 (\opencite{peck04}). This 
possibility has, however, to be confirmed by interferometric measurements. 

As suggested by Sect.\,2, most H$_2$O kilomasers are associated with star formation. There are, however, exceptions. 
Towards M\,51, the kilomaser was found to coincide with the nucleus within 250\,mas ($\sim$10\,pc; \opencite{hagiwara01}). 
A similarly good coincidence, within 1$''$ ($\sim$12\,pc), was also found for the main maser component in the starburst galaxy 
NGC\,253 (\opencite{henkel04}). The nature of these sources that are too weak to be seen at distances well in excess of 10\,Mpc
is still under discussion.

Almost a decade ago, all H$_2$O megamasers were found to be associated with Seyfert 2 and LINER galaxies (\opencite{braatz97}).  
In the meantime, however, with a much larger number of detections, the situation has become more complex. While megamasers
are detected in optically `normal' galaxies (\opencite{greenhill02}), a particularly luminous megamaser was recently found in 
the FRII radio galaxy 3C\,403 (\opencite{tarchi03}). Another megamaser host, NGC\,5506, was identified as a Narrow Line 
Seyfert 1 (\opencite{nagar02}). The detection of a nuclear kilomaser in a second such galaxy, NGC\,4051 (\opencite{hagiwara03}), 
was interpreted in terms of a relatively low inclination of the nuclear disk or torus w.r.t Seyfert 2 galaxies, thus yielding 
relatively low column densities for coherent amplification. Even more recently, H$_2$O was detected in the prototypical Seyfert 
1 NGC\,4151 (\opencite{braatz04}). It is still too early for a detailed interpretation, but it is clear that these detections 
will be essential for a better understanding of the unifying scheme differentiating between type 1 and type 2 AGN. 

Is there a way to predict the presence of H$_2$O megamasers by observations at other wavelengths? This question was recently 
discussed for the prototypical sources NGC\,1068, NGC\,3079 and NGC\,4258 (\opencite{bennert04}). All three galaxies exhibit 
a spatially compact ($\la$1$''$) near infrared core containing dust clouds that are heated by the central engine. This appears 
to be the main hint for the potential presence of accretion disk masers, but these cores are more difficult to detect than the 
masers themselves. Tracers for jet masers at optical and near infrared wavelengths are spectral lines that are split into two 
velocity components.

\section{Future propects}

Though the sensitivity of existing facilities can be slightly enhanced, order(s) of magnitude improvements will only be
possible with the Square Kilometer Array (SKA), complemented by space interferometry to combine optimal sensitivity
with optimal angular resolution. The 22\,GHz line of water vapor bears the prospect to reveal the magnetic field strength 
through its Zeeman pattern. A measurement towards NGC\,4258 yielded a 1$\sigma$ upper limit of 300\,mG for the toroidal
component (\opencite{herrnstein98b}). Probably only the SKA will provide the sensitivity to determine the magnetic field 
in the circumnuclear environment of active galaxies. The same may hold for searches of H$_2$O megamasers at cosmological
redshifts. On the other hand, accounting for the statistical properties of detected maser sources, it appears that existing 
facilities have the potential to drastically enlarge the number of known luminous sources of H$_2$O emission. So far, possibly
only a small percentage of the detectable megamasers have been found (\opencite{peck04}).

\end{article}

\begin{thebibliography}{}
\bibitem[\protect\citeauthoryear{Argon et al.}{2004}]{argon04}
 Argon, A.L., Greenhill, L.J., Reid, M.J. et al. 2004, {\it ApJ\/}, submitted
\bibitem[\protect\citeauthoryear{Bennert et al.}{2004}]{bennert04}
 Bennert, N., Schulz, H. and Henkel, C. 2004, {\it A\&A\/}, {\bf 419}, 127
\bibitem[\protect\citeauthoryear{Braatz et al.}{1996}]{braatz96}
 Braatz, J.A., Wilson, A.S. and Henkel, C. 1996, {\it ApJS\/}, {\bf 106}, 51
\bibitem[\protect\citeauthoryear{Braatz et al.}{1997}]{braatz97}
 Braatz, J.A., Wilson, A.S. and Henkel, C. 1997, {\it ApJS\/}, {\bf 110}, 321
\bibitem[\protect\citeauthoryear{Braatz et al.}{2004}]{braatz04}
 Braatz, J.A., Henkel, C., Greenhill, L.J., Moran, J.M. and Wilson, A.S. 2004, {\it ApJ\/}, in preparation
\bibitem[\protect\citeauthoryear{Bragg et al.}{2000}]{bragg00}
 Bragg, A.E., Greenhill, L.J., Moran, J.M. and Henkel, C. 2000, {\it ApJ\/}, {\bf 535}, 73
\bibitem[\protect\citeauthoryear{Brunthaler et al.}{2002}]{brunthaler02}
 Brunthaler, A., Falcke, H., Reid, M., Greenhill, L.J. and Henkel, C. 2002, {\it Proc. of the 6th European VLBI Symp.},
 eds. Ros et al., MPIfR, Bonn, Germany, p.\,189
\bibitem[\protect\citeauthoryear{Cheung et al.}{1969}]{cheung69}
 Cheung, A.C., Rank, D.M., Townes, C.H., Thornton, D.D. and Welch, W.J. 1969, {\it Nature\/}, {\bf  221}, 626
\bibitem[\protect\citeauthoryear{Churchwell et al.}{1977}]{churchwell77}
 Churchwell, E., Witzel, A., Huchtmeier, W., et al. 1977, {\it A\&A\/}, {\bf 54}, 969 
\bibitem[\protect\citeauthoryear{Claussen et al.}{1984}]{claussen84}
 Claussen, M.J., Heiligman, G.M. and Lo, K.-Y. 1984, {\it Nature}, {\bf 310}, 298
\bibitem[\protect\citeauthoryear{Claussen and Lo}{1986}]{claussen86}
 Claussen, M.J. and Lo, K.-Y. 1986, {\it ApJ\/}, {\bf 308}, 592
\bibitem[\protect\citeauthoryear{Claussen et al.}{1998}]{claussen98}
 Claussen, M.J., Diamond, P.J., Braatz, J.A., Wilson, A.S., and Henkel, C. 1998, {\it ApJ\/}, {\bf 500}, L129
\bibitem[\protect\citeauthoryear{Curran et al.}{1999}]{curran99}
 Curran, S.J., Rydbeck, G., Johansson, L.E.B. and Booth, R.S. 1999, {\it A\&A\/}, {\bf 344}, 767 
\bibitem[\protect\citeauthoryear{Dos Santos and L{\'e}pine}{1979}]{dossantos79}
 Dos Santos, P.M. and L{\'e}pine, J.R.D. 1979, {\it Nature\/}, {\bf 278}, 34
\bibitem[\protect\citeauthoryear{Ferrarese and Merritt}{2000}]{ferrarese00}
 Ferrarese, L. and Merritt, D. 2000, {\it ApJ\/}, {\bf 539}, L9 
\bibitem[\protect\citeauthoryear{Gardner and Whiteoak}{1982}]{gardner82}
 Gardner, F.F. and Whiteoak, J.B. 1982, {\it MNRAS}, {\bf 201}, 13p
\bibitem[\protect\citeauthoryear{Gallimore et al.}{1996}]{gallimore96}
 Gallimore, J.F., Baum, S.A., O'Dea, C.P., Brinks, E. and Pedlar, A. 1996, {\it ApJ}, {\bf 462}, 740
\bibitem[\protect\citeauthoryear{Gallimore et al.}{2001}]{gallimore01}
 Gallimore, J.F., Henkel, C., Baum, S.A., et al. 2001, {\it ApJ}, {\bf 556}, 694 
\bibitem[\protect\citeauthoryear{Greenhill et al.}{1997}]{greenhill97}
 Greenhill, L.J. and Gwinn, C.R. 1997, {\it ApSS\/}, {\bf 248}, 261 
\bibitem[\protect\citeauthoryear{Greenhill et al.}{1993}]{greenhill93}
 Greenhill, L.J., Moran, J.M., Reid, M.J., Menten, K.M. and Hirabayashi, H. 1993, {\it ApJ\/}, {\bf 406}, 482
\bibitem[\protect\citeauthoryear{Greenhill et al.}{1995a}]{greenhill95a}
 Greenhill, L.J., Henkel, C., Becker, R., Wilson, T.L. and Wouterloot, J.G.A. 1995a, {\it A\&A\/}, {\bf 304}, 21
\bibitem[\protect\citeauthoryear{Greenhill et al.}{1995b}]{greenhill95b}
 Greenhill, L.J., Jiang, D.R., Moran, J.M., et al. 1995b, {\it ApJ\/}, {\bf 440}, 619
\bibitem[\protect\citeauthoryear{Greenhill et al.}{1996}]{greenhill96}
 Greenhill, L.J., Gwinn, C.R., Antonucci, R. and Barvainis, R. 1996, {\it ApJ\/}, {\bf 472}, L21 
\bibitem[\protect\citeauthoryear{Greenhill et al.}{2002}]{greenhill02}
 Greenhill, L.J., Ellingsen, S.P., Norris, R.P. et al. 2002, {\it ApJ\/}, {\bf 565}, 836 
\bibitem[\protect\citeauthoryear{Greenhill et al.}{2003}]{greenhill03}
 Greenhill, L.J., Booth, R.S., Ellingsen, S.P., et al. 2003, {\it ApJ\/}, {\bf 590}, 162 
\bibitem[\protect\citeauthoryear{Hagiwara et al.}{2001}]{hagiwara01}
 Hagiwara, Y., Henkel, C., Menten, K.M. and Nakai, N. 2001, {\it ApJ\/} {\bf 560}, L37 
\bibitem[\protect\citeauthoryear{Hagiwara et al.}{2002}]{hagiwara02}
 Hagiwara, Y., Henkel, C., Sherwood, W.A. and Baan, W.A. 2002, {\it A\&A\/} {\bf 387}, L29 
\bibitem[\protect\citeauthoryear{Hagiwara et al.}{2003}]{hagiwara03}
 Hagiwara, Y., Diamond, P.J., Miyoshi, M., Rovilos, E. and Baan, W.A. 2003, {\it MNRAS\/} {\bf 344}, L53 
\bibitem[\protect\citeauthoryear{Haschick and Baan}{1985}]{haschick85}
 Haschick, A.D. and Baan, W.A. 1985, {\it Nature\/} {\bf 314}, 144
\bibitem[\protect\citeauthoryear{Haschick and Baan}{1990}]{haschick90}
 Haschick, A.D. and Baan, W.A. 1990, {\it ApJ\/} {\bf 355}, L23
\bibitem[\protect\citeauthoryear{Haschick et al.}{1994}]{haschick94}
 Haschick, A.D.,  Baan, W.A. and Peng, E.W. 1994, {\it ApJ\/} {\bf 437}, L35
\bibitem[\protect\citeauthoryear{Henkel et al.}{1984}]{henkel84}
 Henkel, C., G{\"u}sten, R., Downes, D., et al. 1984, {\it A\&A\/}, {\bf 141}, L1
\bibitem[\protect\citeauthoryear{Henkel et al.}{2002}]{henkel02}
 Henkel, C., Braatz, J.A., Greenhill, L.J. and Wilson, A.S. 2002, {\it A\&A\/}, {\bf 394}, L23 
\bibitem[\protect\citeauthoryear{Henkel et al.}{2004}]{henkel04}
 Henkel, C., Tarchi, A., Menten, K.M. and Peck, A.B. 2004, {\it A\&A\/}, {\bf 414}, 117 
\bibitem[\protect\citeauthoryear{Herrnstein et al.}{1998a}]{herrnstein98a}
 Herrnstein, J.R., Greenhill, L.J., Moran, J.M., et al. 1998a, {\it ApJ\/}, {\bf 497}, L69
\bibitem[\protect\citeauthoryear{Herrnstein et al.}{1998b}]{herrnstein98b}
 Herrnstein, J.R., Moran, J.M., Greenhill, L.J., Blackman, E.G. and Diamond, P.J. 1998b, {\it ApJ\/}, {\bf 508}, 243
\bibitem[\protect\citeauthoryear{Herrnstein et al.}{1999}]{herrnstein99}
 Herrnstein, J.R., Moran, J.M., Greenhill, L.J., et al. 1999, {\it Nature\/}, {\bf 400}, 539
\bibitem[\protect\citeauthoryear{Ishihara et al.}{2001}]{ishihara01}
 Ishihara, Y., Nakai, N., Iyomoto, N., et al. 2001, {\it PASJ\/}, {\bf 53}, 215 
\bibitem[\protect\citeauthoryear{Kondratko et al.}{2004}]{kondratko04}
 Kondratko, P.T., Greenhill, L.J. and Moran, J.M. 2004, {\it ApJ\/}, in preparation
\bibitem[\protect\citeauthoryear{Miyoshi et al.}{1995}]{miyoshi95}
 Miyoshi, M., Moran, J.M., Herrnstein, J.R., et al. 1995, {\it Nature\/}, {\bf 373}, 127
\bibitem[\protect\citeauthoryear{Moran et al.}{1999}]{moran99}
 Moran, J.M., Greenhill, L.J., Herrnstein, J.R. 1999, {\it JApA\/}, {\bf 20}, 165
\bibitem[\protect\citeauthoryear{Nagar et al.}{2002}]{nagar02}
 Nagar, N.M., Oliva, E., Marconi, A. and Maiolino, R. 2002, {\it A\&A\/}, {\bf 391}, L21
\bibitem[\protect\citeauthoryear{Nakai et al.}{1993}]{nakai93}
 Nakai, N., Inoue, M. and Miyoshi, M. 1993, {\it Nature\/}, {\bf 361}, 45
\bibitem[\protect\citeauthoryear{Nakai et al.}{1995}]{nakai95}
 Nakai, N., Inoue, M., Miyazawa, K., Miyoshi, M. and Hall, P. 1995, {\it PASJ\/}, {\bf 47}, 771 
\bibitem[\protect\citeauthoryear{Neufeld and Maloney}{1995}]{neufeld95}
 Neufeld, D.A. and Maloney, P.R. 1995, {\it ApJ\/}, {\bf 447}, L17 
\bibitem[\protect\citeauthoryear{Peck et al.}{2003}]{peck03}
 Peck, A.B., Henkel, C., Ulvestad, J.S., et al. 2003, {\it ApJ\/}, {\bf 590}, 149 
\bibitem[\protect\citeauthoryear{Peck et al.}{2004}]{peck04}
 Peck, A.B., Tarchi, A., Henkel, C., et al. 2004, {\it ApJ\/}, in preparation
\bibitem[\protect\citeauthoryear{Schulz and Henkel}{2003}]{schulz03}
 Schulz, H. and Henkel, C. 2003, {\it A\&A\/}, {\bf 400}, 41 
\bibitem[\protect\citeauthoryear{Tarchi et al.}{2002}]{tarchi02}
 Tarchi, A., Henkel, C., Peck, A.B. and Menten, K.M. 2002, {\it A\&A\/}, {\bf 389}, L39
\bibitem[\protect\citeauthoryear{Tarchi et al.}{2003}]{tarchi03}
 Tarchi, A., Henkel, C., Chiaberge, M. and Menten, K.M. 2003, {\it A\&A\/}, {\bf 407}, L33
\bibitem[\protect\citeauthoryear{Trotter et al.}{1998}]{trotter98}
 Trotter, A.S., Greenhill, L.J., Moran, J.M., et al. 1998, {\it ApJ\/}, {\bf 495}, 740
\bibitem[\protect\citeauthoryear{Veilleux and Bland-Hawthorn}{1997}]{veilleux97}
 Veilleux, S. and Bland-Hawthorn, J. 1997, {\it ApJ\/}, {\bf 479}, L105
\end{thebibliography}
\end{document}